\documentclass[prl,onecolumn,reprint,aps,superscriptaddress,longbibliography]{revtex4-1}
\usepackage{graphicx}%
\usepackage{dcolumn}%
\usepackage{bm}%
\usepackage{color}
\usepackage{multirow}
\usepackage[normalem]{ulem}
\usepackage{amsmath}
\usepackage{enumerate}
\usepackage{amsfonts}
\usepackage{epsfig}
\usepackage{graphicx}
\usepackage{cancel}
\usepackage[caption=false]{subfig}
\usepackage{floatrow}
\usepackage{subfig}
\newcommand{\be}{\begin{equation}}
\newcommand{\ee}{\end{equation}}
\newcommand{\bea}{\begin{eqnarray}}
\newcommand{\eea}{\end{eqnarray}}
\definecolor{green}{rgb}{0.0, 0.44, 0.0}
\definecolor{red}{rgb}{1.0, 0.13, 0.32}
\definecolor{blue}{rgb}{0.06, 0.2, 0.65}
\definecolor{darkgreen}{rgb}{0,0.5,0}
\definecolor{darkblue}{rgb}{0,0,0.6}
\definecolor{purple}{rgb}{0.4,.2,0.7}
\definecolor{magenta}{rgb}{1.0,0.0,1.0}
\usepackage[colorlinks=true,citecolor=darkgreen,linkcolor=purple,urlcolor=purple]{hyperref}

\def\le{\left}
\def\ri{\right}
\def\lg{\ell_{\mathrm{grad}}}
\def\noi{\eta}
\begin{document}

%\title{Collective gradient-sensing through supervised and unsupervised learnings}
%\title{Optimizing gradient-sensing capability of swarming agents through collective learning}
%\title{Training swarming agents to collectively sense gradients through reinforcement}
%\title{Reinforcing a collective mind to efficiently sense gradient}
%\title{Optimizing collective fieldtaxis through reinforcement learning}
\title{Optimizing collective fieldtaxis of swarming agents through reinforcement learning}

\author{Glenn Palmer}
\affiliation{Department of Chemistry, Duke University, Durham, North Carolina 27708, USA}

\author{Sho Yaida}
\email{sho.yaida@duke.edu}
\affiliation{Department of Chemistry, Duke University, Durham, North Carolina 27708, USA}

\begin{abstract}
Swarming of animal groups enthralls scientists in fields ranging from biology to physics to engineering. Complex swarming patterns often arise from simple interactions between individuals to the benefit of the collective whole. The existence and success of swarming, however, nontrivially depend on microscopic parameters governing the interactions. Here we show that a machine-learning technique can be employed to tune these underlying parameters and optimize the resulting performance. As a concrete example, we take an active matter model inspired by schools of golden shiners, which collectively conduct phototaxis. The problem of optimizing the phototaxis capability is then mapped to that of maximizing benefits in a continuum-armed bandit game. The latter problem accepts a simple reinforcement-learning algorithm, which can tune the continuous parameters of the model. This result suggests the utility of machine-learning methodology in swarm-robotics applications.
\end{abstract}

\maketitle

\emph{Introduction--} 
The ubiquity of collective swarming has fascinated scientists from all walks of life, including biologists~\cite{Allee31,PE99,CDFSTB01,KR02}, physicists~\cite{VZ12,MJRLPMA13,CG14,Popkin16}, and roboticists~\cite{OFL04,HLVRCZF11,MMCMEHPS12,Bouffanais16}. For instance, birds flock~\cite{Heppner74,Heppner97}, fish school~\cite{PP80,Partridge82}, fireflies sync~\cite{Strogatz04}, neurons think~\cite{Hopfield82}, and ants build exotic structures such as bridges~\cite{RLPKCG15,GKWG17} and towers~\cite{WPN14}. Certain collective behaviors benefit animal communities in foraging for foods and sheltering from predators, whereas others can be entertaining but sometimes dangerous, such as Mexican waves at sporting events~\cite{FHV02} and mosh pits at heavy metal concerts~\cite{SBSC13,JSM00}. Such behaviors, though once mused to be manifestations of telekinetic effects~\cite{Selous31}, can arise from simple local interaction rules among swarming agents~\cite{Aoki82,Reynolds87,VCBCS95,CKJRF02}. Even with underlying mechanisms demystified, it is still tempting to view swarming agents as forming a collective mind~\cite{Couzin07}. This perspective prompts a natural follow-up question: can this mind learn? More practically, can we optimize -- or even reverse engineer -- collective behaviors of swarming robots to our liking through an arsenal of machine-learning techniques~\cite{Mitchell97,SB98,RN03,KF09,Murphy12,GBC16}?

Collective foraging provides an instructive and fruitful testing ground in that it offers an obvious process to be optimized. Namely, preference is given to the fast and reliable tracking of food sources (or equivalently of dark regions to hide from predators). Some biological~\cite{Engelmann1881,Pfeffer1888} and synthetic~\cite{HBKSV07} organisms find foods through chemotaxis at an individual level by sensing gradient in chemical densities (also see Ref.~\cite{VVS07}). A recent study of golden shiners, \textit{Notemigonus crysoleucas}, however, revealed that schools of fish can
%could
navigate to regions of lower light intensity through collective phototaxis even if individuals
%were
are incapable of detecting light-field gradient~\cite{BTIFC13}. Various mechanisms for collective gradient sensing have since been proposed, such as field-dependent contact inhibition of locomotion interaction for chemotaxis~\cite{CZLR16} and long-range force transmission for durotaxis~\cite{SCEELVNGMRT16} -- a fundamental process in healing wounds.

The discovery of naturally occurring collective fieldtaxis has subsequently triggered an effort to build robotic swarms with the same gradient-sensing mechanism~\cite{WCZ12,Shinerbot}. These robots could ultimately become indispensable in applications ranging from sensing chemical pollutants to disaster rescue missions. The efficiency of collective swarming, however, critically hinges on the microscopic parameters governing local interactions between agents. Thus, in contrast to traditional approaches aimed at inferring interaction rules that dictate actual animal groups~\cite{KTIHC11,BCGMSVW12,CG14,CGGMPTW14,NZB17}, in robotic applications the focus must be put on tuning system parameters to achieve optimal outcomes within given engineering constraints. This circles back to our original question. In this letter, we propose that machine-learning methodology can be adapted to efficiently improve capabilities of swarming robots. We in particular demonstrate the efficacy of reinforcement learning in optimizing the collective fieldtaxis performance of a simple model. This work paves the way for assembling more complex swarming patterns through optimization in high-dimensional parameter spaces.

\begin{figure*}
\vspace{-0.3in}
\centerline{
\subfloat[\ \ \ \ \ \ \ \ \ \ \ \ \ \ \ \ ]{\includegraphics[width=0.39\textwidth]{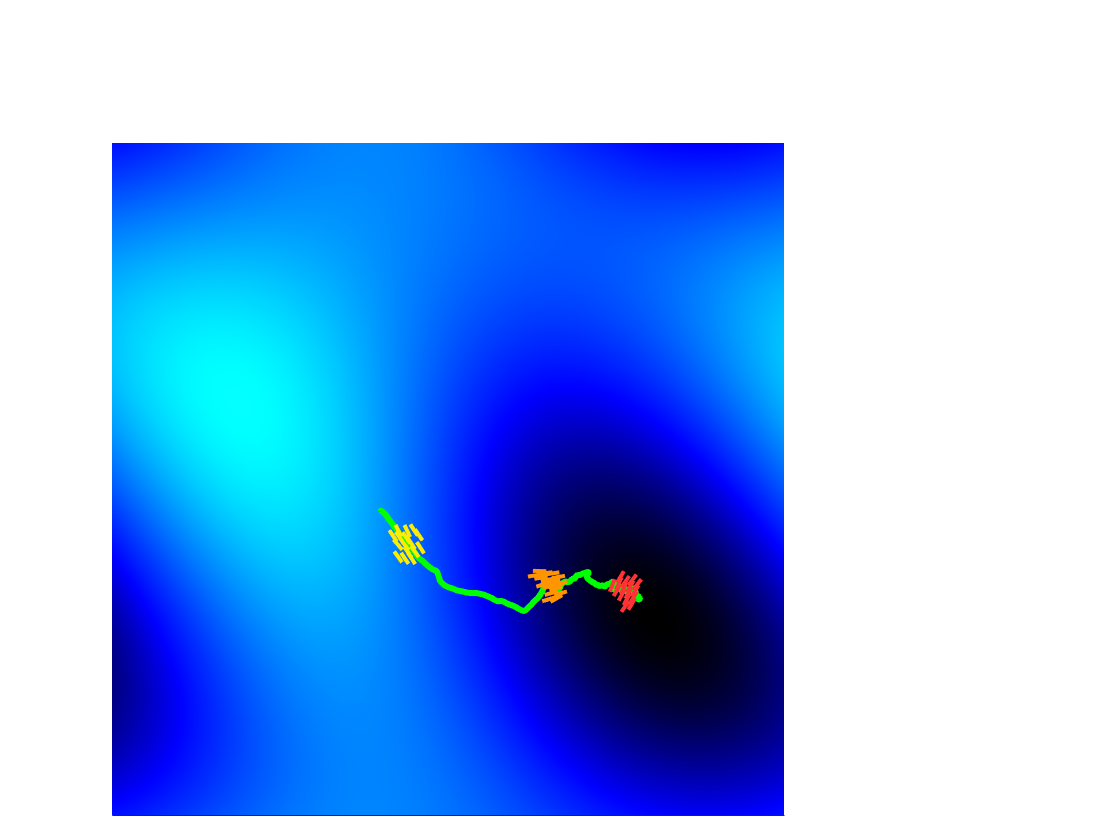}}\quad
\hspace{-0.8in}
\subfloat[\ \ \ \ \ \ \ \ \ \ \ \ \ \ \ \ ]{\includegraphics[width=0.39\textwidth]{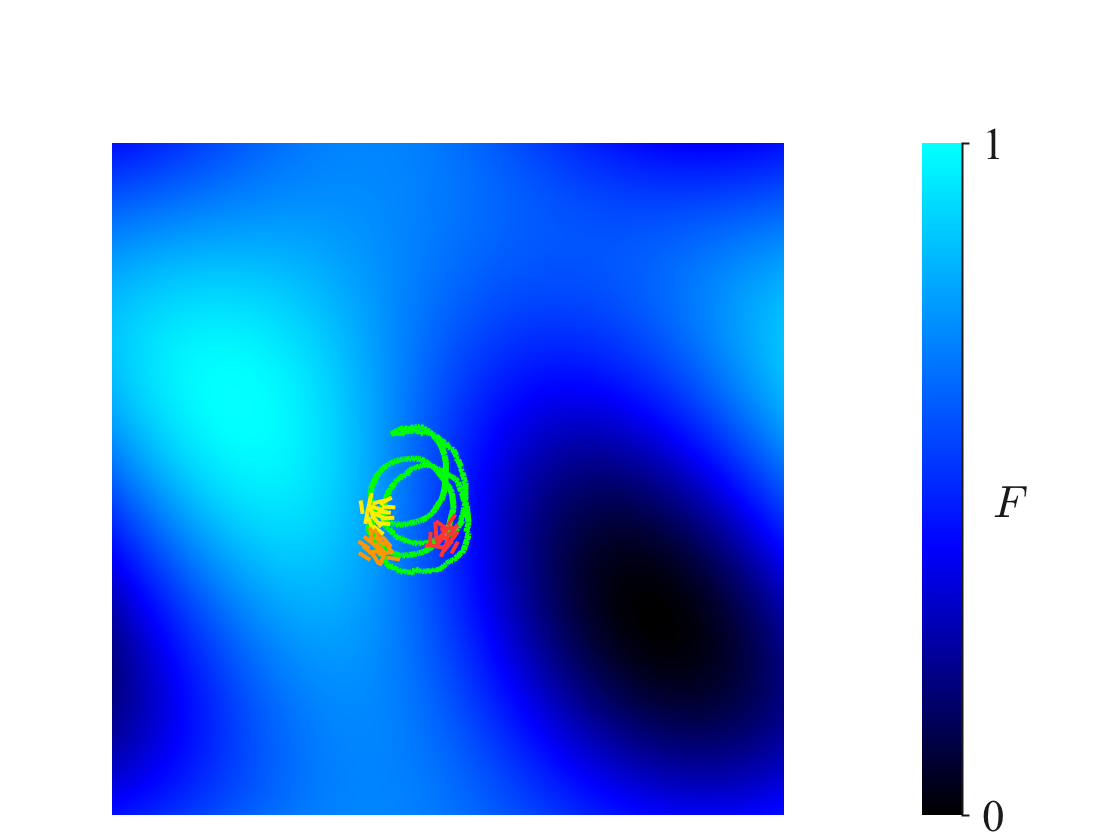}}\quad
\hspace{-0.1in}
\subfloat[\ \ \ ]{\includegraphics[width=0.33\textwidth]{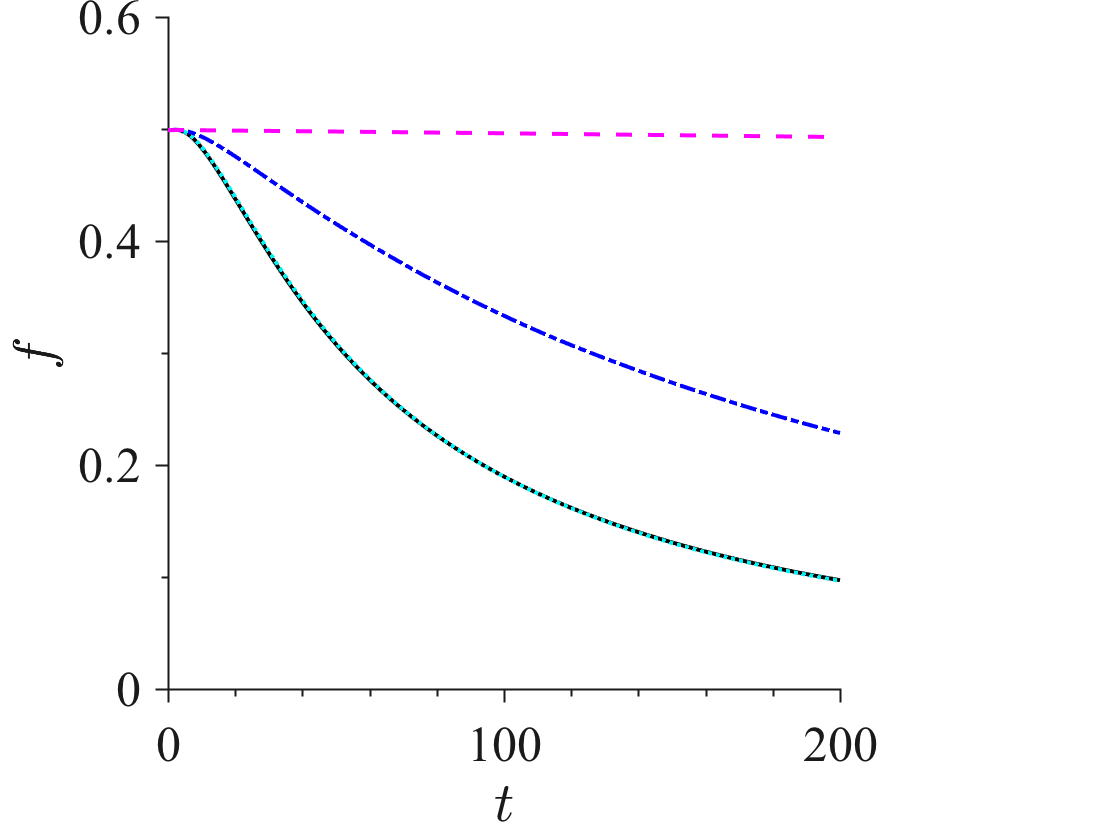}}
}
\caption{Dynamical trajectories of $N=16$ swarming agents in the periodic box of linear size $L=100$. (a) A typical trajectory for a trained group with $(r_{\mathrm{o}},r_{\mathrm{a}})=(2.06,2.20)$ after time $t=10$ (yellow), $100$ (orange), and $1000$ (red). The green line depicts the center-of-mass trajectory of the group from $t=0$ to $10000$. The group finds the minimum without difficulty. (b) A similar trajectory of a demoralized group with $(r_{\mathrm{o}},r_{\mathrm{a}})=(1.70,2.00)$ results in the group wandering. (c) The time dependence of the field intensity perceived by a group, $f$, defined in Eq.~\eqref{fave}. The result is averaged over $10^6$ local environments of light fields~\cite{footnote_laziness}.
At $t=t_{\star}=20$, the trained group (black-solid) performs better than the demoralized one (magenta-dashed), a pre-training (see Fig.~\ref{bandit}) one with $(r_{\mathrm{o}},r_{\mathrm{a}})=(1.95,2.05)$ (blue-dash-dotted), and a suboptimal one with $(r_{\mathrm{o}},r_{\mathrm{a}})=(2.10,2.25)$ (cyan-dotted). A close inspection reveals that the suboptimal group starts to slightly win over the optimal one after $t\gtrsim45$, reflecting that the definition of optimality depends on the time scale of interest.
%Errorbars are of the order of $10^{-4}$ and hence invisible.
}
\label{trajectories}
\end{figure*}

\emph{Model--}
A group exhibits collective gradient sensing when (i) it flocks with a coherent center-of-mass velocity and (ii) the coherent velocity accelerates toward the direction of the gradient in the field. As is well known, coherent flocking takes place in systems of self-propelled agents interacting through repulsion at short range $r_{\mathrm{r}}$, orientation aligning at intermediate range $r_{\mathrm{o}}$, and attraction at long range $r_{\mathrm{a}}$ for some choices of these parameters. In addition, as has been realized in making minimal models of golden shiners~\cite{BTIFC13}, the acceleration of the coherent velocity can be induced by the magnitude dependence of velocity on the local field intensity, with relatively slow motion in preferred regions. These considerations naturally lead to the following active matter model.

Let $\le\{\mathbf{x}_i(t),\mathbf{v}_i(t)\ri\}_{i=1,\ldots,N}$ denote positions and velocities of $N$ agents in the two-dimensional periodic box of linear size $L$, where $t$ denotes time. At location $\mathbf{X}$, agents feel a field intensity $F(\mathbf{X})$ ranging from zero to one, with zero indicating preferred regions, be they shades, food sources, or chemical spills. With each time step $\Delta t$, positions and velocities are updated as
\bea
\mathbf{x}_i(t+\Delta t)&=&\mathbf{x}_i(t)+\frac{1}{2}\le[\mathbf{v}_i(t)+\mathbf{v}_i(t+\Delta t)\ri]\, \\
\mathbf{v}_i(t+\Delta t)&=&\le[v_{\mathrm{max}}F\le(\mathbf{x}_i(t)\ri)\ri] R_{\mathrm{noise}}\hat{\mathbf{d}}_i(t)\,  ,
\eea
where $v_{\mathrm{max}}$ denotes the maximum velocity and a matrix
\be
R_{\mathrm{noise}}=\begin{bmatrix}
\cos(\theta_{\mathrm{noise}}) & -\sin(\theta_{\mathrm{noise}}) \\ \sin(\theta_{\mathrm{noise}}) & \cos(\theta_{\mathrm{noise}})\, 
\end{bmatrix}
\ee
represents noise inherent in information processing, modeled by drawing 
$\theta_{\mathrm{noise}}$ uniformly from $[-\noi, \noi]$ with the noise level specified by $\noi$. The normalized directional vector $\hat{\mathbf{d}}_i(t)\equiv\mathbf{d}_i(t)/\|\mathbf{d}_i(t)\|$ is determined via the following rules: (i) if there are any neighbors within the zone of repulsion $r_{\mathrm{r}}$, then
\be
\mathbf{d}_i(t)=-\sum_{\substack{j\ne i \\ \|\mathbf{x}_{j}(t)-\mathbf{x}_{i}(t)\|<r_{\mathrm{r}}}} \frac{\mathbf{x}_{j}(t)-\mathbf{x}_{i}(t)}{\|\mathbf{x}_{j}(t)-\mathbf{x}_{i}(t)\|}\, ;
\ee
(ii) if there are no neighbors in the zone of repulsion but some within the zone of orientation/attraction $r_{\mathrm{o,a}}$, then
\bea
\mathbf{d}_i(t)&=&\le[\sum_{\substack{j\ne i \\ r_{\mathrm{r}}\leq\|\mathbf{x}_{j}(t)-\mathbf{x}_{i}(t)\|<r_{\mathrm{o}}}} \frac{\mathbf{v}_{j}(t)}{\|\mathbf{v}_{j}(t)\|}\ri]\, \\
&&+\le[\sum_{\substack{j\ne i \\ r_{\mathrm{o}}\leq\|\mathbf{x}_{j}(t)-\mathbf{x}_{i}(t)\|<r_{\mathrm{a}}}} \frac{\mathbf{x}_{j}(t)-\mathbf{x}_{i}(t)}{\|\mathbf{x}_{j}(t)-\mathbf{x}_{i}(t)\|}\ri]\, ;\nonumber
\eea
and (iii) if there are no neighbors within the zone of attraction, then $\mathbf{d}_i(t)=\mathbf{v}_i(t)$. Here ranges are restricted to the parameter subspace in which $r_{\mathrm{r}}\leq r_{\mathrm{o}}\leq r_{\mathrm{a}}$.

Lengths can be measured in the unit of the repulsion range $r_{\mathrm{r}}$ and time in the unit of $t_0\equiv r_{\mathrm{r}}/v_{\mathrm{max}}$, both set to unity henceforth. The box size, $L=100$, is set large enough to avoid boundary effects, the time step is chosen to be $\Delta t=0.01$ to comply with observed short information-processing time~\cite{CG14}, and the noise level is set to be small, $\noi=0.1\sqrt{\Delta t}$, as the effect of noise is not of primary interest here~\cite{footnote_noise}. In addition to the number of agents $N$ and the gradient length scale introduced by a light field $F$, this essentially leaves two parameters governing the behavior of the model, $r_{\mathrm{o}}$ and $r_{\mathrm{a}}$.

Typical trajectories of the model are depicted in Fig.~\ref{trajectories}. When the microscopic parameters are near optimal, the group readily finds a minimum of the field, as expected. For other parameters, by contrast, the group ceases to move or simply wanders, dissipating energy through random swarming motion without coherent direction. The success and efficiency of the collective fieldtaxis thus crucially depend on the microscopic parameters, $(r_{\mathrm{o}},r_{\mathrm{a}})$.

The optimal efficiency can be found by a brute-force parameter search if there is only one continuous parameter characterizing the model or if parameters are discrete (see, e.g., Ref.~\cite{MCMOL13}). As the number of parameters increases, however, such an approach quickly becomes intractable, with the requisite computational time roughly increasing as $\frac{1}{\epsilon^p}$ where $\epsilon$ is the desired accuracy and $p$ is the number of continuous parameters. For generic systems with multiple parameters, an alternative approach is thus imperative. We now demonstrate that a simple variant of reinforcement algorithms~\cite{SB98}, which have been applied to optimize behavior at an individual level~\cite{RCSV16,CGCB17}, can be effective at a collective level.\\

\emph{Training Algorithm--}
Our training algorithm repeats the following protocol: cast a static field, initialize agents at $t=0$, record the reward obtained up to time $t=t_{\star}$, and adjust parameters through reinforcement. These training sessions are indexed by $\alpha=1,\ldots,N_{\rm train}$.

Specifically, a light field is generated by defining
\be
\tilde{F}^{(\alpha)}(\mathbf{X})=\sum_{\mathbf{k}} \le[A^{(\alpha)}_{\mathbf{k}}\cos(\mathbf{k}\cdot\mathbf{X})+B^{(\alpha)}_{\mathbf{k}}\sin(\mathbf{k}\cdot\mathbf{X})\ri]\, ,
\ee
shifted and rescaled so that it ranges precisely from zero to one. Here the sum over wavevectors, $\mathbf{k}=(k_x,k_y)$, runs through $k_x,k_y=0,\frac{2\pi}{L},\ldots,\frac{2\pi n_{\mathrm{max}}}{L}$ and amplitudes $A^{(\alpha)}_{\mathbf{k}}$ and  $B^{(\alpha)}_{\mathbf{k}}$ are independently drawn from a normal distribution~\cite{footnote_Gauss}. Agents are then initialized to positions $\mathbf{x}^{(\alpha)}_{i}(0)$ uniformly within a circle of radius $R=\sqrt{\frac{N}{\pi}}$ with a randomly chosen center. Velocities are in turn initialized to be $\mathbf{v}^{(\alpha)}_{i}(0)=\le[-v_{\mathrm{max}} F \frac{\nabla F}{\|\nabla F\|}\ri]\le(\mathbf{X}=\mathbf{x}^{(\alpha)}_{i}(0)\ri)\,$.

In order to promote the decay of the average field intensity perceived by agents,
\be\label{fave}
f(t)\equiv \frac{1}{N}\sum_{i=1}^{N} F(\mathbf{x}_i(t))\, ,
\ee
the reward is defined as
\be
Q\equiv\max\{0, f(0)-f(t_{\star})\}\, 
\ee
with a fixed terminal time $t_{\star}$.
In other words, if $f(t)$ decreases over time duration $t_{\star}$, the magnitude of its decrease is the reward; otherwise, there is no reward.

For each training, a pair of trial parameter values $\le(r^{(\alpha)}_{\mathrm{o}},r^{(\alpha)}_{\mathrm{a}}\ri)$ is chosen from the continuum of allowed values, and a reward is given probabilistically based on that choice. Our goal is to identify the pair of parameters, $\le(r^{\star}_{\mathrm{o}},r^{\star}_{\mathrm{a}}\ri)$, that maximizes the average reward, which is equivalent to the continuum-armed bandit problem~\cite{Agrawal95}. This generalization of the classical multi-armed bandit problem~\cite{Robbins52,SB98} has a simple learning algorithm that performs stochastic gradient ascent to shift the parameters toward optimality as the training proceeds~\cite{Kleinberg05,FKM05}.
Namely, at the $\alpha$-th training session with a light field and an initial configuration drawn as above, the rewards are evaluated at three nearby points in the parameter space: $Q_{0}$ at $\le(r^{(\alpha)}_{\mathrm{o}},r^{(\alpha)}_{\mathrm{a}}\ri)$, $Q_{1}$ at $\le(r^{(\alpha)}_{\mathrm{o}}-\delta,r^{(\alpha)}_{\mathrm{a}}\ri)$, and $Q_{2}$ at $\le(r^{(\alpha)}_{\mathrm{o}},r^{(\alpha)}_{\mathrm{a}}+\delta\ri)$ with deviation $\delta\equiv (\alpha+1)^{-1/4}$. The parameters are then updated as
\bea
r^{(\alpha+1)}_{\mathrm{o}}&=&r^{(\alpha)}_{\mathrm{o}}+\gamma(Q_0-Q_1)/\delta\, \\
r^{(\alpha+1)}_{\mathrm{a}}&=&r^{(\alpha)}_{\mathrm{a}}+\gamma(Q_2-Q_0)/\delta\, 
\eea
with $\gamma\equiv(\alpha+1)^{-3/4}$~\cite{footnote_bandit}. Over many iterations, these updates obtain an ascent in the landscape of the average reward function $Q(r_{\mathrm{o}}, r_{\mathrm{a}})$ without evaluating such a function for every pair of parameters.

\begin{figure}
\centerline{
\includegraphics[width=0.84\textwidth]{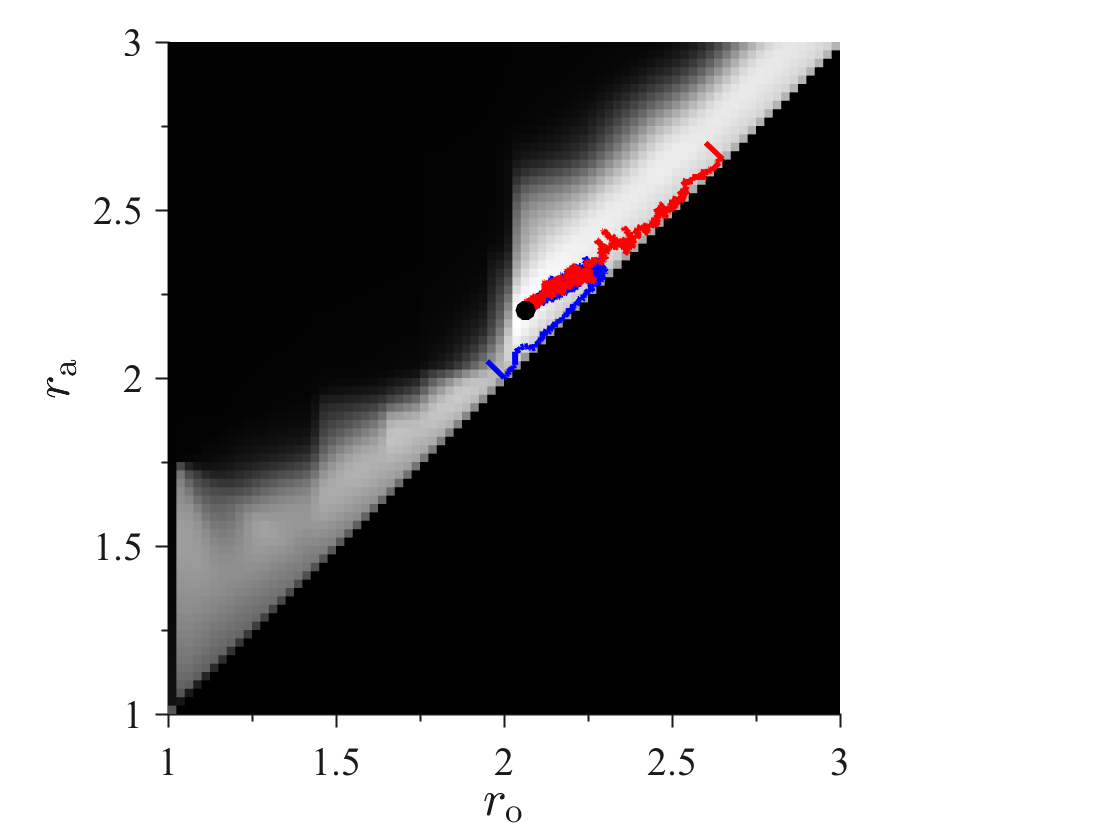}
}
\vspace{-2.28in}
\centerline{
\hspace{+4.4in}\includegraphics[width=0.71\textwidth]{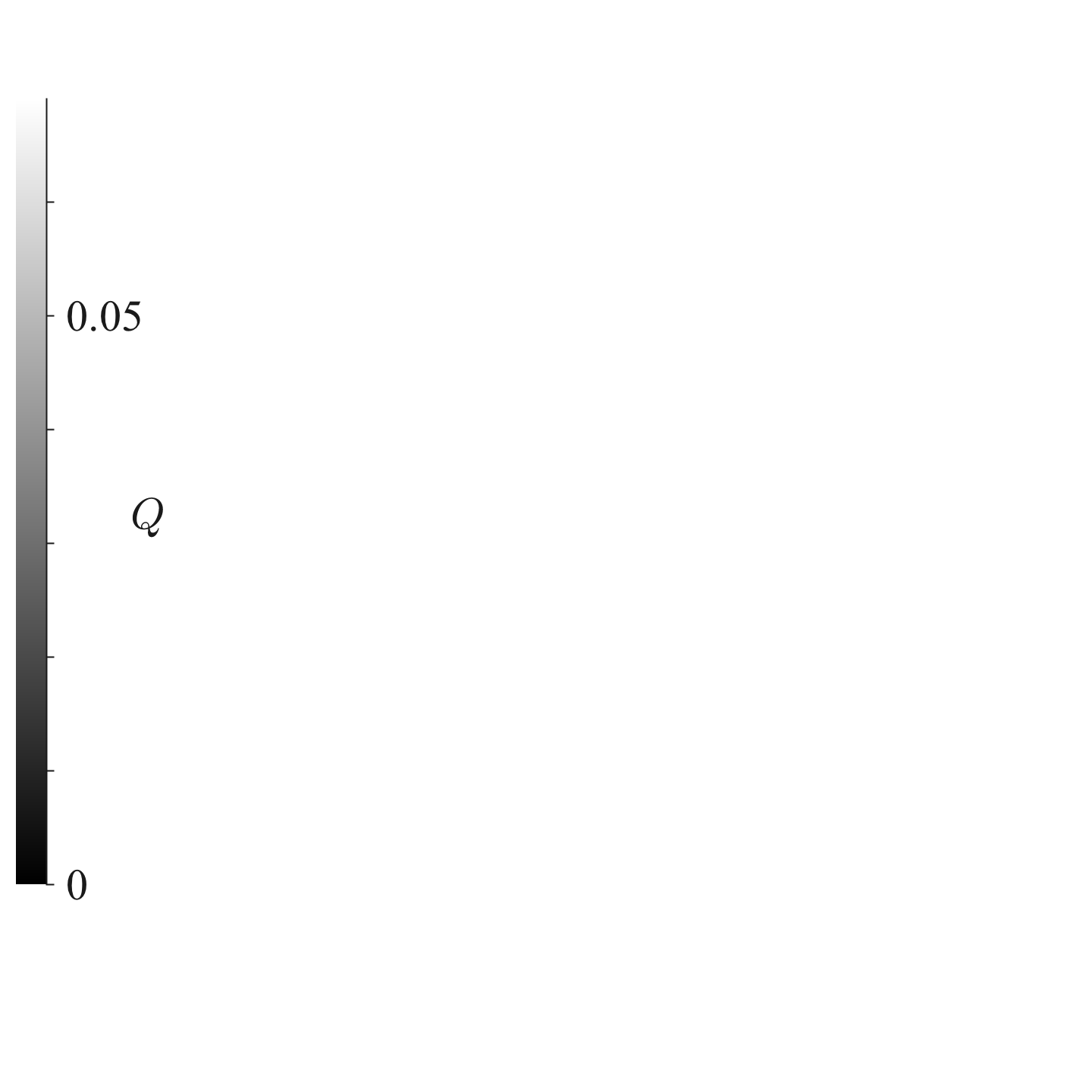}
}
\vspace{-0.23in}
\caption{Training trajectories of model parameter values in parameter space.
Trajectories with distinct initial guesses, $\le(r^{(\alpha)}_{\mathrm{o}},r^{(\alpha)}_{\mathrm{a}}\ri)\Big|_{\alpha=1}=(1.95,2.05)$ (blue line) and $(2.6,2.7)$ (red line), converge to the same optimal point $\le(r^{\star}_{\mathrm{o}},r^{\star}_{\mathrm{a}}\ri)\approx(2.06,2.20)$ (black dot) of the average reward function $Q(r_{\mathrm{o}}, r_{\mathrm{a}})$ (displayed as the background color and obtained through the brute-force parameter scan). The learning algorithm scales better upon increasing the number of model parameters and, even for the two-parameter case, requires much less computational time than the brute-force search.
}
\label{bandit}
\end{figure}

In Fig.~\ref{bandit} we depict training trajectories of the parameters with the terminal time $t_{\star}=20$ for $N=16$ and $n_{\mathrm{max}}=1$, which sets the gradient scale to be roughly of order $\lg\equiv\frac{L}{2\pi n_{\mathrm{max}}}\approx 16$.  The training has been carried out over $N_{\mathrm{train}}=10^6$ sessions~\cite{footnote_laziness}. Also depicted is the outcome of the brute-force parameter search, obtained with grid spacing $0.025$ in the range $1\leq r_{\mathrm{o}}\leq r_{\mathrm{a}}\leq3$, at each point averaged over the $10^5$ realizations of local environments~\cite{footnote_laziness}. The resultant error bars in $Q(r_{\mathrm{o}}, r_{\mathrm{a}})$ are of order $10^{-4}$, barely sufficient to locate the optimum at this grid resolution. Bandit-algorithm trajectories converge to the optimum as long as initial guesses are chosen reasonably well, but with much less computational time and greater accuracy than the brute-force search. The bandit algorithm is further expected to scale better as the number of model parameters $p$ increases, with power-law dependence on $p$~\cite{Kleinberg05,FKM05} in contrast to the exponential dependence of the brute-force search. The algorithm thus generalizes well.

\begin{figure}
\centerline{
\quad\quad
\sidesubfloat[]{\hspace{-0.05in}\includegraphics[width=0.55\textwidth]{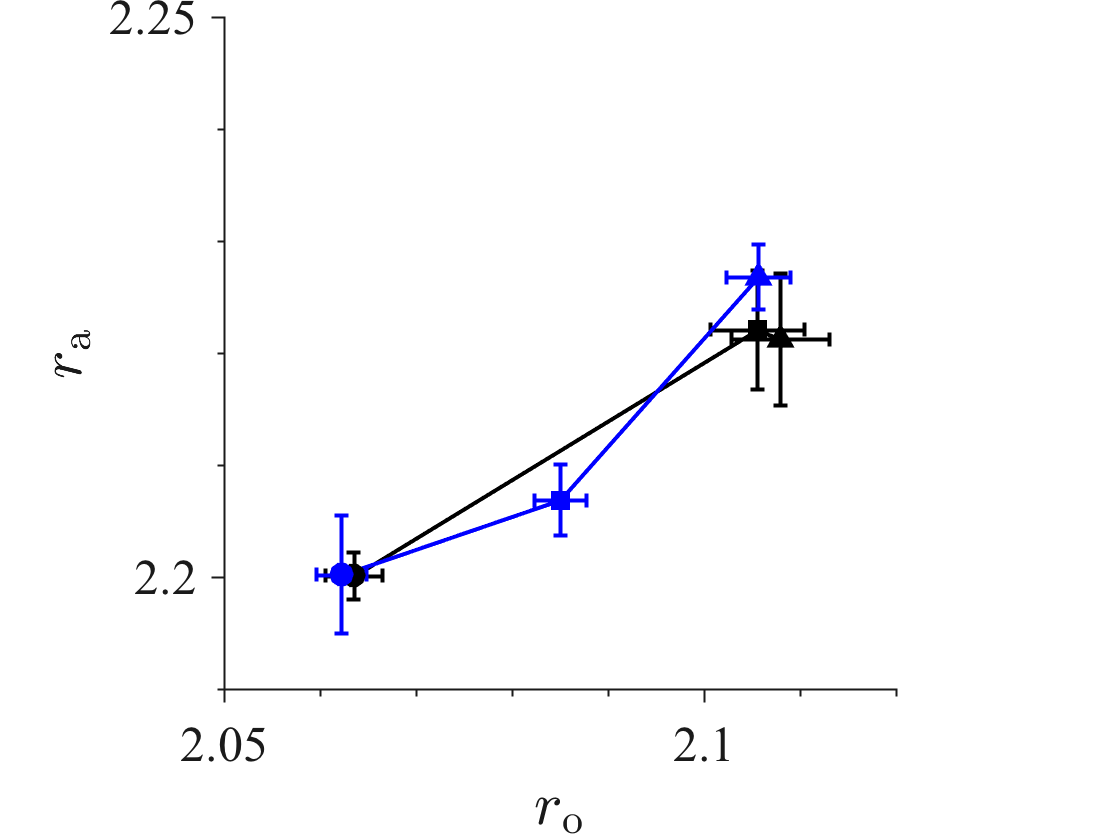}}\quad
\hspace{-0.5in}
\sidesubfloat[]{\hspace{-0.05in}\includegraphics[width=0.55\textwidth]{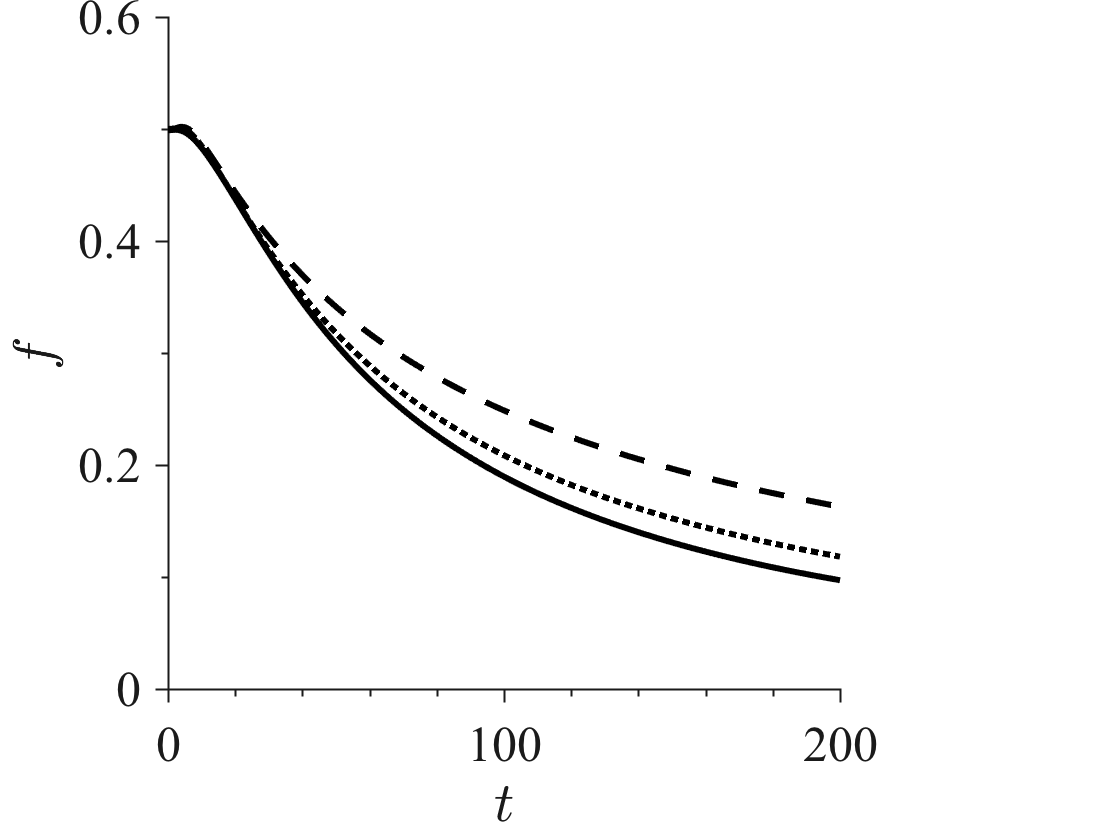}}
}
\caption{Effects of variations in environmental parameters. (a) The positions of optimal parameters obtained by the bandit algorithm, averaged over $10$ distinct initial guesses, for $N=16$ (circle), $32$ (square), and $64$ (triangle), with the mild gradient $n_{\mathrm{max}}=1$ (black) and shaper gradient $n_{\mathrm{max}}=2$ (blue). Lines are guide for the eyes. (b) The field intensity perceived by a group, $f$, as a function of time, $t$, for the parameter $(r_{\mathrm{o}},r_{\mathrm{a}})=(2.06,2.20)$ found to be optimal for $N=16$. The different curves correspond to $N=16$ (solid), $32$ (dotted), and $64$ (dashed).}
\label{environmental}
\end{figure}

It is important to keep in mind that the choices of $t_{\star}$, $N$, and $n_{\mathrm{max}}$ affect the reward and hence the exact position of optimality. For instance, while the trained group maximizes decrease in the field intensity over time $t_{\star}$, some nearby parameters may -- and indeed do -- yield slightly better performance at other time scales [Fig.~\ref{trajectories}(c)]. Similarly, the optimal parameters shift as $N$ is varied, even though small-group training extends reasonably well upon scaling up group size  (Fig.~\ref{environmental}).
%If wished or more natural, one can incorporate the variability of these parameters into the optimization process by varying $N$ and $n_{\mathrm{max}}$ in the training set.
%Finally, realistic fields may dynamically shift. The trained group successfully tracks low-intensity regions of fields (movie online).

\emph{Conclusion--} In this paper, we have implemented a learning algorithm to optimize a specific reward for effective collective gradient sensing. Unlike some evolutionary algorithms that result in compromised optimality due to idiocy of selfish agents~\cite{TBC11,HRHBTC15,Bennati16,TLCL15,KMTHC14}, convergence to optimality is guaranteed for a reasonable initial guess. More broadly, this simple example suggests that machine-learning insights can be brought to bear on the optimization problems that arise in applications of swarm robotics.

We have further found that the optimality achieved depends on the choice of reward. A simplistic choice of reward and algorithm has been made here, especially in that there is no history dependence. If memory or history are taken into account, it might be possible to employ more elaborate and efficient reinforcement algorithms~\cite{Watkins89,WD92,SB98,RCSV16,CGCB17}. If, instead, exemplary motion trajectories to be mimicked are given, then we can use apprenticeship~\cite{AN04} or supervised learning, minimizing the deviation from such motions. In particular, we may be able to combine multiple collective swarming behaviors that belong to distinct species. This could lead us to the boundary of learning capability, as in self-assembly constructions~\cite{MZBL15,MZB15} and community detection~\cite{DKMZ11,ZK16,Moore17}. In this regard, it is clear that the learning capability depends on the model used. Perhaps one can construct a general-purpose active matter model with many parameters that allows for fast and more generic learning, in the same vein as the deep-neural network that has been so successful in beating games such as Atari~\cite{MKSRVBGRFOPBSAKKWLH15} and Go~\cite{SHMGSVSAPLDGNKSLLKGH16}.

\begin{acknowledgments}
We thank Gordon J.~Berman, Patrick Charbonneau, Iain D.~Couzin, Daniel I.~Goldman, Liesbeth M.~C.~Janssen,
Stanley Kok, Joyjit Kundu, Arvind Murugan, and Daniel M.~Sussman for stimulating discussions and suggestions.
We also acknowledge the Duke Compute Cluster for computational times.
This work was supported by grants from the National Science Foundation's Research Triangle Materials Research Science and Engineering Center (No.~DMR-1121107) and from the Simons Foundation (No.~454937, Patrick Charbonneau).
%Data relevant to this work have been archived and can be accessed at http://dx.doi.org/xx.xxxx/XxxXxXxX.
\end{acknowledgments}

\bibliography{CL}

\begin{thebibliography}{75}
\bibitem [{foo({\natexlab{a}})}]{footnote_laziness}%
  \BibitemOpen
  \href@noop \ \bibinfo {note} {Since the system size
  $L=100$ is large enough, in order to sample over many local environments, we
  have prepared $10^4$ independent light fields and recycled over them with
  different initial positions of agents.}\BibitemShut {Stop}%
\bibitem [{foo({\natexlab{b}})}]{footnote_noise}%
  \BibitemOpen
  \href@noop \ \bibinfo {note} {When the noise level is
  set to zero, the entire group sometimes comes to a complete
  halt.}\BibitemShut {Stop}%
\bibitem [{foo({\natexlab{c}})}]{footnote_Gauss}%
  \BibitemOpen
  \href@noop \ \bibinfo {note} {The result does not
  depend on the mean and variance of the Gaussian distribution upon
  rescaling.}\BibitemShut {Stop}%
\bibitem [{foo({\natexlab{d}})}]{footnote_bandit}%
  \BibitemOpen
  \href@noop \ \bibinfo {note} {When the update takes
  the parameters outside the region $r_{\mathrm{o}}\leq r_{\mathrm{a}}$, they
  are further shifted to
  $(\frac{r_{\mathrm{o}}+r_{\mathrm{a}}}{2},\frac{r_{\mathrm{o}}+r_{\mathrm{a}}}{2})$.}\BibitemShut
  {Stop}%
\end{thebibliography}

\end{document}